\documentclass{article}
\usepackage{spconf}
\usepackage{graphicx,cite,amssymb,amsmath,psfrag,subfigure}
\usepackage{cite}
\usepackage{mathrsfs}
\usepackage[displaymath,mathlines]{lineno}
\usepackage{color}
\usepackage{tabulary}
\usepackage{multirow}
\usepackage{cases}
\usepackage{stfloats}
\usepackage{url}
\usepackage[subfigure]{tocloft}
\usepackage{import}

\newtheorem{remark}{Remark}

\def\bD{{\boldsymbol{D}}}

\def\bg{{\boldsymbol{g}}}
\def\bG{{\boldsymbol{G}}}

\def\bI{{\boldsymbol{I}}}

\def\bw{{\boldsymbol{w}}}

\def\bx{{\boldsymbol{x}}}

\def\by{{\boldsymbol{y}}}

\newcommand{\pp}[1]{{\left( #1 \right)}}

\newcommand{\snorm}[1]{{ \left\Vert #1 \right\Vert^2 }}
\newcommand{\fsnorm}[1]{{ \left\Vert #1 \right\Vert^2_F }}

\newcommand{\E}[1]{{ \mbox{E}\left[ #1 \right] }}

\newcommand{\var}[1]{{ \mathbb{V}\mathrm{ar}\left( #1 \right) }}

\def\sumk{{{\sum_{k=-\infty}^\infty}}}

\def\sumk{{{\sum_{k=1}^{K}}}}

\def\lambdamax{{{\sigma_{\mbox{\tiny max}}}}}
\def\lambdamin{{{\sigma_{\mbox{\tiny min}}}}}

\def\CN{{{CN}}}

\begin{document}
\title{Aspects of Favorable Propagation in Massive MIMO\vspace{-0.5cm}}

\name{Hien Quoc Ngo$^*$, Erik G. Larsson$*$, Thomas L. Marzetta$^\dagger$ %
   \thanks{The work of H.~Q.\ Ngo and E.~G.\ Larsson was supported  in part
by the Swedish Research Council (VR), the Swedish Foundation for
Strategic Research (SSF), and ELLIIT.}%
}
\address{%
   $^*$ Department of Electrical Engineering (ISY),
Link\"{o}ping University, 581 83 Link\"{o}ping, Sweden\\
   $^\dagger$ Bell Laboratories, Alcatel-Lucent
Murray Hill, NJ 07974, USA\\\vspace{-0.5cm} }


\maketitle


\begin{abstract}
Favorable propagation, defined as mutual orthogonality among the
vector-valued channels to the terminals, is one of the key
properties of the radio channel that is exploited in Massive MIMO.
 However, there has been little work that studies   this topic in detail.
 In this paper, we  first  show that
favorable propagation offers the most desirable scenario in terms
of maximizing the sum-capacity. One useful proxy for whether propagation is
 favorable or not is the channel condition number. However, this proxy is not good for
 the case where the norms of the channel vectors may not be
equal. For this case, to evaluate how favorable the propagation
offered by the channel is, we propose a ``distance from favorable
propagation'' measure, which is the gap between the sum-capacity
and the maximum capacity obtained under favorable propagation.
Secondly, we examine how favorable the channels can be for two
extreme scenarios: i.i.d.~Rayleigh fading and uniform random
line-of-sight (UR-LoS). Both environments offer (nearly) favorable
propagation. Furthermore, to analyze the UR-LoS model, we propose
an urns-and-balls model. This model is simple and explains   the
singular value spread characteristic of the UR-LoS model well.


\end{abstract}


\section{Introduction}
Recently, there has been a great deal of interest in massive
multiple-input multiple-output (MIMO) systems where a base
station equipped with a few hundred antennas simultaneously serves
several tens of terminals
\cite{Mar:10:WCOM,HBD:13:JSAC,LTEM:13:CM}. Such systems can
deliver all the attractive benefits of traditional MIMO, but at a
much larger scale. More precisely, massive MIMO systems can
provide high throughput, communication reliability, and high power
efficiency  with linear processing \cite{NLM:13:TCOM2}.

One of the key assumptions exploited by massive MIMO   is that the
channel vectors between the base station and the terminals should
be nearly orthogonal. This is called \emph{favorable propagation}.
With favorable propagation, linear processing can achieve optimal
performance. More explicitly, on the uplink, with a simple linear
detector  such as the matched filter, noise and interference can
be canceled out. On the downlink, with linear beamforming
techniques, the base station can simultaneously beamform multiple
data streams to multiple terminals without causing mutual
interference. Favorable propagation of massive MIMO was discussed
in the papers \cite{RPLLMET:11:SPM,NLM:13:TCOM2}. There,   the
condition number of the channel matrix was used as a proxy to
evaluate how favorable the channel is. These papers only
considered the case that the channels are i.i.d. Rayleigh fading.
However, in practice, owing to the fact that the terminals have
different locations, the norms of the channels are not identical.
As we will see here, in this case,  the condition number is not a
good proxy for whether or not we have favorable propagation.

In this paper, we investigate the favorable propagation condition
of different channels. We first show that under favorable
propagation, we maximize the sum-capacity under a power
constraint. When the channel vectors are i.i.d., the singular
value spread is a useful proxy to evaluate how  favorable the
propagation environment is. However, when the channel vectors have
different norms, this is not so. We also ask whether or not
practical scenarios will lead to  favorable propagation.  To this
end, we consider two extreme scenarios: i.i.d.~Rayleigh fading and
uniform random line-of-sight (UR-LoS). We show that both scenarios
offer substantially favorable propagation. We also propose a
simple urns-and-balls model to analyze the UR-LoS case. For the
sake of the argument, we  will consider the uplink of a
single-cell system.

\section{Single-Cell System Model}
Consider the uplink of a single-cell system where $K$
single-antenna terminals independently and simultaneously transmit
data to the base station. The base station has $M$ antennas and
all $K$ terminals share the same time-frequency resource. If the
$K$ terminals simultaneously transmit the $K$ symbols
$x_1,\ldots,x_K$, where $\E{|x_k|^2}=1$, then the $M\times 1$
received vector at the base station is
\begin{align}\label{eq:ulrx2}
\by = \sqrt{\rho} \sum_{k=1}^K \bg_k x_k + \bw =  \sqrt{\rho} \bG
\bx + \bw,
\end{align}
where $\bx=[x_1,\ldots,x_K]^T$, $\bG=[\bg_1,\ldots,\bg_K]$,
$\bg_k\in \mathbb{C}^{M\times 1}$ is the channel vector between
the base station and the $k$th terminal, and $\bw$ is a noise
vector. We assume that the elements of $\bw$ are i.i.d.\ $\CN(0,1)$
RVs. With this assumption, ${\rho}$ has the interpretation of
normalized ``transmit'' signal-to-noise ratio (SNR). The channel
vector $\bg_k$ incorporates the effects of large-scale fading and small-scale
fading. More precisely, the $m$th element of $\bg_k$ is modeled
as:
\begin{align}\label{eq:defgkm}
g_k^m = \sqrt{\beta_k} h_k^m,\qquad k=1,\ldots,K,\quad
m=1,\ldots,M,
\end{align}
where $h_k^m$ is the small-scale fading and $\beta_k$ represents
the large-scale fading which depends on $k$ but not on
$m$.\vspace{-0.3cm}

\section{Preliminaries of Favorable Propagation}\vspace{-0.3cm}
In favorable propagation, we can obtain optimal performance with
simple linear processing techniques. To have favorable
propagation, the channel vectors $\{\bg_k\}$, $k=1,\ldots, K$,
should be pairwisely orthogonal. More precisely, we say that the
channel offers \emph{favorable propagation} if
\begin{align}\label{eq:deffp}
 \bg_{i}^H\bg_{j} =
\left\{%
\begin{array}{l}
  0,\quad  i, j=1,\ldots,K,\quad  i\neq  j \\
  \snorm{\bg_k} \neq  0, \quad k=1,\ldots,K. \\
\end{array}%
\right.
 \end{align}
In practice,  the condition (\ref{eq:deffp}) will never be exactly
satisfied, but   (\ref{eq:deffp}) can be approximately achieved.
For this case, we say that the channel offers \emph{approximately
favorable propagation}. Also, under some assumptions on the
propagation environment, when $M$ grows large and $k\neq j$, it
holds that
\begin{align}
\frac{1}{M} \bg_{k}^H\bg_{j} \to & 0, \quad
M\to\infty.\label{eq:deffpas}
 \end{align}
For this case, we say that the channel offers \emph{asymptotically
favorable propagation}.

The favorable propagation condition \eqref{eq:deffp}  does not
offer only the optimal performance with linear processing but also
represents the most desirable scenario from the perspective of
maximizing the information rate. See  the following section.
\vspace{-0.3cm}
\subsection{Favorable Propagation and Capacity}
Consider the system model \eqref{eq:ulrx2}. We assume that the
base station knows the channel $\bG$. The sum-capacity is given by
\begin{align}\label{eq:macrate1}
C  = \log_2 \left| \bI + \rho\bG^H \bG \right|.
\end{align}
Next, we will show that, subject to a constraint on $\bG$, under
favorable propagation conditions \eqref{eq:deffp},   $C$
  achieves its largest possible value.
Firstly, we assume $\{\snorm{\bg_k}\}$ are given. For this case,
by using the Hadamard inequality, we have
\begin{align}\label{eq:Rfp1}
C  = & \log_2 \left| \bI + \rho\bG^H \bG \right| \le \log_2 \pp{
\prod_{k=1}^K [\bI + \rho\bG^H \bG ]_{k,k}} \nonumber
\\
= & \sumk \!\log_2\! \pp{ \! [\bI \!+\! \rho\bG^H \bG ]_{k,k}\!}
\!=\! \sumk\! \log_2 \!\pp{   \!1 \!+\! \rho \snorm{\bg_k}\! }.
\end{align}
We can see that the equality of \eqref{eq:Rfp1} holds if and only
if $\bG^H\bG$ is diagonal, so that (\ref{eq:deffp}) is satisfied.
This means that, given a constraint on $\{\snorm{\bg_k}\}$, the
channel propagation with the condition (\ref{eq:deffp}) provides
the maximum sum-capacity.

Secondly, we consider a more relaxed constraint on the channel
$\bG$: constraint $\fsnorm{\bG}$ instead of $\{\snorm{\bg_k}\}$.
From (\ref{eq:Rfp1}), by using Jensen's inequality, we get
\begin{align}\label{eq:Rfp2}
C  \le & \sumk  \log_2 \pp{   1 + \rho \snorm{\bg_k} }
\!=\! K\cdot \frac{1}{K} \sumk \! \log_2\! \pp{   \!1 + \rho \snorm{\bg_k} \!}  \nonumber \\
\le & K \!\log_2\!\!\pp{\! \!1\!+\! \frac{\rho}{K}\sumk
\snorm{\bg_k}\!}
 \!=\!  K \!\log_2\! \pp{\!1 + \frac{\rho}{K} \fsnorm{\bG}\!},
\end{align}
where the equality in the first  step holds when (\ref{eq:deffp})
satisfied, and the equality in the second step holds when all
$\snorm{\bg_k}$ are equal. So, for this case, $C$ is maximized if
(\ref{eq:deffp}) holds and $\{\bg_k\}$ have the same norm. The
constraint on $\bG$ that results in (\ref{eq:Rfp2}) is more
relaxed than the constraint on $\bG$ that results in
(\ref{eq:Rfp1}), but the bound in (\ref{eq:Rfp2}) is only  tight
if all $\{\bg_k\}$ have the same norm.


\subsection{Measures of Favorable Propagation}
Clearly, to check whether the channel can offer favorable
propagation or not, we can check directly the condition
\eqref{eq:deffp}  or \eqref{eq:deffpas}. However, to do this, we
have to check all $(K-1)K/2$ possible pairs. This has
computational complexity. Other simple methods to measure whether
the channel offers favorable propagation is to  consider the
condition number, or the \emph{distance from favorable
propagation} (to be defined shortly). These measures will be
discussed in more detail in the following subsections.

\subsubsection[Condition Number]{Condition Number}
Under the favorable propagation condition \eqref{eq:deffp}, we
have
\begin{align}\label{eq:CN 1}
\bG^H\bG =\mathrm{Diag}\{\snorm{\bg_1}, \cdot\cdot\cdot,
\snorm{\bg_K}\}.
\end{align}
We can see that if $\{\bg_k\}$ have the same norm, the condition
number of the Gramian matrix $\bG^H \bG$ is equal to 1:
\begin{align}
\lambdamax/\lambdamin =1,
\end{align}
where $\lambdamax$ and $\lambdamin$ are the maximal and minimal
singular values of $\bG^H\bG$.

Similarly, if the channel offers asymptotically favorable
propagation, then we have
\begin{align}\label{eq:CN 1b}
\bG^H\bG \to \bD, \quad M\to\infty,
\end{align}
where $\bD$ is a diagonal matrix whose $k$th diagonal element is
$\beta_k$. So, if all $\{\beta_k\}$ are equal, then the condition
number is asymptotically equal to $1$.

Therefore, when the channel vectors have  the same norm (the large
scale fading coefficients are equal), we can use the condition
number to determine how favorable the channel propagation is.
Since the condition number  is simple to evaluate, it has been
used as a measure of how favorable the propagation offered by the
channel $\bG$ is, in the literature. However, it has two
 drawbacks: i) it only has a sound operational meaning when all $\{\bg_k\}$ have the same norm or all
$\{\beta_k\}$ are equal; and ii) it disregards all other singular
values than $\lambdamin$ and $\lambdamax$.

\vspace{-0.3cm}
\subsubsection{Distance from Favorable Propagation}\vspace{-0.2cm}
As discussed above, when $\{\bg_k\}$ have different norms or
$\{\beta_k\}$ are different, we cannot use the condition number to
measure how favorable the  propagation is. For this case, we
propose to use the \emph{distance from favorable propagation}
which is defined as the relative gap between the capacity $C$
obtained by this propagation and the upper bound in
\eqref{eq:Rfp1}:
\begin{align}
\Delta_C \triangleq \!\frac{\sumk \! \log_2\!\pp{\! 1\! +\! \rho
\snorm{\bg_k}} \!-\! \log_2\! \left|\! \bI \!+\! \rho\bG^H \bG
\!\right|}{\log_2 \left| \bI + \rho\bG^H \bG \right|}.
\end{align}
The distance from favorable propagation represents how far from
favorable propagation the channel is. Of course, when
$\Delta_C$=0, we have favorable propagation.


\vspace{-0.3cm}
\section{Favorable Propagation: Rayleigh Fading and Line-of-Sight Channels}\vspace{-0.2cm}
One of the key properties of Massive MIMO systems is that the
channel under some conditions can offer asymptotically favorable
propagation. The basic question is, under what conditions is the
channel favorable? A more general question is what practical
scenarios result in favorable propagation. In practice, the
channel properties depends a lot on the propagation environment as
well as the antenna configurations. Therefore, there are varieties
of channel models such as Rayleigh fading, Rician, finite
dimensional channels, keyhole channels, LoS, etc. In this section,
we will consider two particular channel models:  independent
Rayleigh fading and uniform random line-of-sight (UR-LoS). These
channels represent very different physical scenarios. We will
study how favorable these channels are and compare the singular
value spread. For simplicity, we set $\beta_k=1$ for all $k$ in
this section.

\vspace{-0.3cm}
\subsection{Independent Rayleigh Fading}
Consider the channel model \eqref{eq:defgkm} where $\{h_k^m\}$ are
i.i.d.\ $\CN(0,1)$ RVs. By using the law of large numbers, we have
 \begin{align}\label{eq:iidrayfp}
 \frac{1}{M} \snorm{\bg_k} & \to 1, \qquad M\to\infty, \qquad\mbox{and} \\
 \frac{1}{M} \bg_k^H \bg_{j} & \to 0, \qquad M\to\infty,\qquad k\neq j,\label{eq:iidrayfpb}
 \end{align}
so we have asymptotically favorable propagation.

In practice, $M$ is large but finite.
Equations~\eqref{eq:iidrayfp}--\eqref{eq:iidrayfpb} show the
asymptotic results when $M\to\infty$ goes to infinity. But, they
do not give an account for how close to favorable propagation the
channel is when $M$ is finite. To study this fact, we consider
 $\var{\frac{1}{M} \bg_k^H \bg_{j}}$. For finite $M$, we have
 \begin{align}\label{eq:iidrayfp1}
\var{ \frac{1}{M}\bg_k^H \bg_{j}} = \frac{1}{M}.
\end{align}
We can see that, $\frac{1}{M} \bg_k^H \bg_{j}$ is concentrated
around $0$ (for $k\neq j$ or $1$ (for $k=j$) with the variance is
proportional to $1/M$.

Furthermore, in Massive MIMO,   the quantity
$\left|\bg_k^H \bg_{j}\right|^2$ is of particular interest. For
example, with matched filtering, the power of the desired signal
is proportional to $\left\|\bg_k\right\|^4$, while the power of
the interference is proportional to $\left|\bg_k^H
\bg_{j}\right|^2$, where $k\neq j$. For $k\neq j$, we have that
\begin{align}\label{eq:iidrayfp2}
 \frac{1}{M^2}  |\bg_k^H \bg_{j}|^2 &\to 0 , \\
\var{ \frac{1}{M^2} |\bg_k^H \bg_{j}|^2} & = \frac{M+2}{M^3}
\approx \frac{1}{M^2}.\label{eq:iidrayfp2b}
 \end{align}
 Equation \eqref{eq:iidrayfp2} shows the convergence of the random quantities $\{\left|\bg_k^H \bg_{j}\right|^2\}$ when $M\to\infty$ which represents the asymptotical
favorable propagation of the channel, and \eqref{eq:iidrayfp2b}
shows the speed of the convergence.

\subsection{Uniform Random  Line-of-Sight}\label{sect:UR LoS}
We consider a scenario with only free space non-fading line of
sight propagation between the base station and the terminals. We
assume that the antenna array is  uniform and  linear  with
antenna spacing $d$. Then in the far-field regime, the channel
vector  $\bg_k$ can be modelled as:
\begin{align}\label{eq:urlosfp0}
\bg_k \!=\! \begin{bmatrix} 1 \!\!&\!\! e^{-i2\pi
\frac{d}{\lambda} \sin(\theta_k)} \!\!&\! \!\cdots\! \!& e^{-i2\pi
\left(M-1\right) \frac{d}{\lambda} \sin(\theta_k)}\!
\end{bmatrix}^T,
\end{align}
where $\theta_k$ is the arrival angle from the $k$th terminal
measured relative to the array boresight, and $\lambda$ is the
carrier wavelength.

For any fixed and distinct angles $\{\theta_k\}$, it is
straightforward to show that
 \begin{align}\label{eq:urlosfp}
 \frac{1}{M} \snorm{\bg_k} & = 1, ~ \mbox{and} ~
 \frac{1}{M} \bg_k^H \bg_{j}  \to 0,  M\to\infty,  k\neq
 j,
 \end{align}
so we have asymptotically favorable propagation.

Now assume that  the $K$ angles $\{\theta_k\}$ are randomly and
independently chosen such that $\sin(\theta_k)$ is uniformly
distributed in $[-1,1]$. We refer to this case as \emph{uniformly
random line-of-sight}. In this case, and if additionally
$d=\lambda/2$, then
 \begin{align}\label{eq:urlosfp1var}
\var{ \frac{1}{M}\bg_k^H \bg_{j}} = \frac{1}{M}-\frac{1}{M^2}.
\end{align}

Comparing \eqref{eq:iidrayfp1} and \eqref{eq:urlosfp1var}, we
see that the inner products between different channel vectors
$\bg_k$ and $\bg_{j}$ converge to zero with the same rate for both
i.i.d.\ Rayleigh fading and in UR-LoS. Interestingly, for finite
$M$, the convergence is slightly faster in the UR-LoS case.

Now consider the quantity $\left|\bg_k^H \bg_{j}\right|^2$. For the
UR-LoS scenario, with $k\neq j$, we have
\begin{align}\label{eq:LoSrayfp2}
 \frac{1}{M^2}  |\bg_k^H \bg_{j}|^2 &\to 0 , \\
\var{ \frac{1}{M^2} |\bg_k^H \bg_{j}|^2} & =
\frac{(M\!-\!1)M(2M\!-\!1)}{3M^4} \!\approx\! \frac{2}{3M}.
\label{eq:LoSrayfp2b}
 \end{align}

We next compare \eqref{eq:iidrayfp2b} and \eqref{eq:LoSrayfp2b}.
While the convergence of the inner products between $\bg_k$ and
$\bg_{j}$ has the same rate in both i.i.d.\ Rayleigh fading and
UR-LoS, the convergence of $\left|\bg_k^H \bg_{j}\right|^2$ is
considerably slower in the UR-LoS case.

\subsection{Urns-and-Balls Model for UR-LoS}
\begin{figure}[t!]
\centerline{\includegraphics[width=0.45\textwidth]{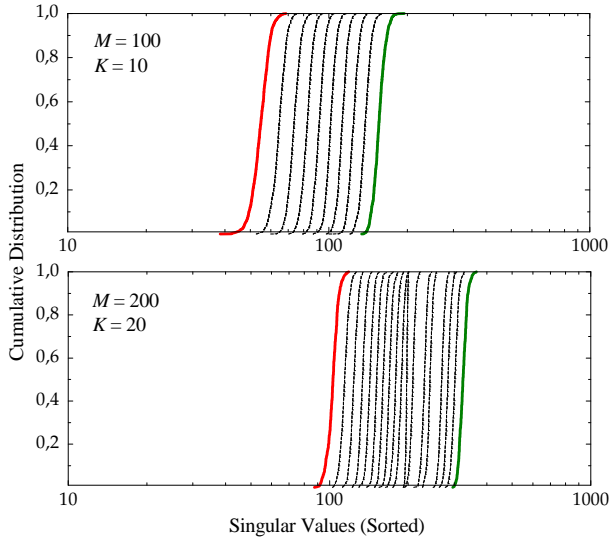}}
\caption[Singular values in i.i.d.~Rayleigh fading and
UR-LoS.]{Singular values of $\bG^H\bG$ for i.i.d.\ Rayleigh
fading. Here, $(M=100,K=10)$ and
$(M=200,K=20)$.\label{fig:fp2b_1}}
\end{figure}

In Section~\ref{sect:UR LoS}, we assumed that angles
$\{\theta_k\}$ are fixed and distinct regardless of $M$. With this
assumption, we have asymptotically favorable propagation. However,
if there exist $\{\theta_k\}$ and $\{\theta_j\}$ such that
$\sin(\theta_k)-\sin(\theta_j)$ is in the  order of
$1/M$, then we cannot have favorable propagation. To see this,
assume that $\sin(\theta_k)-\sin(\theta_j) = 1/M$. Then
 \begin{align}
 \frac{1}{M} \bg_k^H \bg_{j} &=\frac{1}{M}\frac{1-e^{i\pi\left(\sin(\theta_k)-\sin(\theta_j)\right)M}}{1-e^{i\pi\left(\sin(\theta_k)-\sin(\theta_j)\right)}}
 = \frac{1}{M}\frac{1-e^{i\pi}}{1-e^{i\pi/M}}\nonumber
 \\
 &\to \frac{2i}{\pi} \neq 0, \quad M\to\infty.
 \end{align}

\begin{figure}[t!]
\centerline{\includegraphics[width=0.45\textwidth]{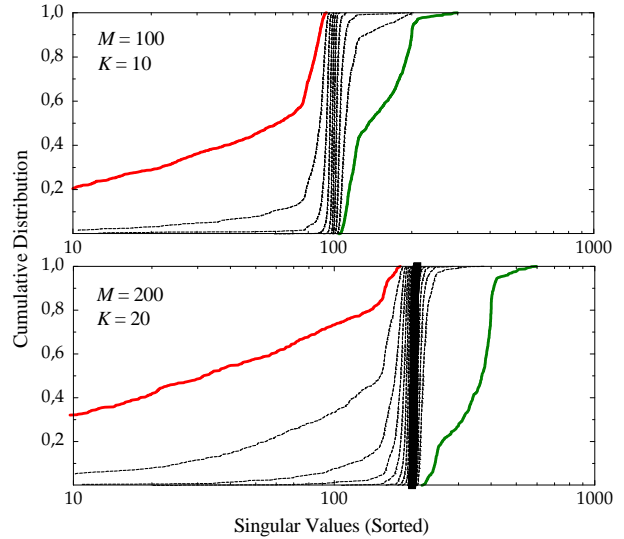}}
\caption{Same as Figure~\ref{fig:fp2b_1}, but for
UR-LoS.\label{fig:fp2b_2}}
\end{figure}

In practice, $M$ is finite. If  the number of terminals $K$ is in
order of tens, then  the probability that there exist
$\{\theta_k\}$ and $\{\theta_j\}$ such that
$\sin(\theta_k)-\sin(\theta_j)\leq 1/M$  cannot be neglected.
This makes the channel unfavorable. This insight can be confirmed
by the following examples. Let consider the singular values of the
Gramian matrix $\bG^H\bG$. Figures~\ref{fig:fp2b_1} and
\ref{fig:fp2b_2} show the cumulative probabilities of the singular
values of $\bG^H\bG$ for i.i.d. Raleigh fading and UR-LoS
channels, respectively. We can see that in i.i.d. Rayleigh fading,
the singular values are uniformly spread out between $\lambdamin$
and $\lambdamax$. However, for UR-LoS, two (for the case of
$M=100, K=10$) or three (for the case of $M=200, K=20$) of the
singular values are very small with a high probability. However,
the rest are highly concentrated around their median. Therefore,
in order to have favorable propagation, we must drop some terminals
from service. In the above examples, we must drop two terminals for
the case $M=100, K=10$ or three terminals for the case $M=200,
K=20$, with high probability.

To quantify approximately how many terminals that have to be
dropped from service so that we have favorable propagation with
high probability in the UR-LoS case, we propose to use the
following simplified model. The base station array can create $M$
orthogonal beams with the angles $\left\{\theta_m \right\}$:
\begin{align}\label{eq:unifangle1}
\sin\left(\theta_m\right) = -1+\frac{2m-1}{M}, \qquad m=1,2, ...,
M.
\end{align}
Suppose that each one of the $K$ terminals is randomly and
independently assigned to one of the $M$ beams given in
\eqref{eq:unifangle1}. To guarantee the channel is favorable, each
beam must contain at most one terminal. Therefore, if there are
two or more terminals in the same beam, all but one of those
terminals must be dropped from service. Let $N_0$, $M-K\leq N_0 <
M$, be the number of beams which have no terminal. Then, the
number of terminals that have to be dropped from service is
\begin{align}\label{eq:unifangle1cb}
N_{\tiny\mbox{drop}}=N_0 - \left(M-K\right).
\end{align}
By using a standard combinatorial result given in
\cite[Eq.~(2.4)]{WF:57:Book1}, we obtain the probability that $n$
terminals, $0 \leq n < K$, are dropped as follows:
\begin{align}\label{eq:urlosdrop1abc}
    &P\left(\!N_{\tiny\mbox{drop}}\!=\! n \!\right)
    =
    P\left(\!N_0 \!-\!\left(\!M-K\!\right)\!= \!n \right)
    =
    P\left(\!N_0\!=\!n\!+\!M\!-\!K \!\right)
    \nonumber\\
    &\!=\!
    \binom {M} {n\!+\!M\!-\!K}
    \nonumber
    \\
    &\hspace{0.8cm}\times
    \sum_{k=1}^{K-n}\!
        \left(\!-1 \!\right)^k \binom{K\!-\!n}{k}
        \left(\!1\!-\!\frac{n\!+\!M\!-\!K\!+\!k}{M} \!\right)^K.
\end{align}
Therefore, the average number of terminals that must be dropped
from service is
\begin{align}\label{eq:urlosdrop1ab}
    \bar{N}_{\tiny\mbox{drop}}=\sum_{n=1}^{K-1} n
    P\left(N_{\tiny\mbox{drop}}= n \right).
\end{align}

\begin{remark}
The result obtained in this subsection yields an important
insight: for Rayleigh fading, terminal selection schemes will not
substantially improve the performance since the singular values
are uniformly spread out. By contrast, in UR-LoS, by dropping some
selected terminals from service, we can improve the worst-user
performance significantly.
\end{remark}

\begin{figure}[t!]
\centerline{\includegraphics[width=0.49\textwidth]{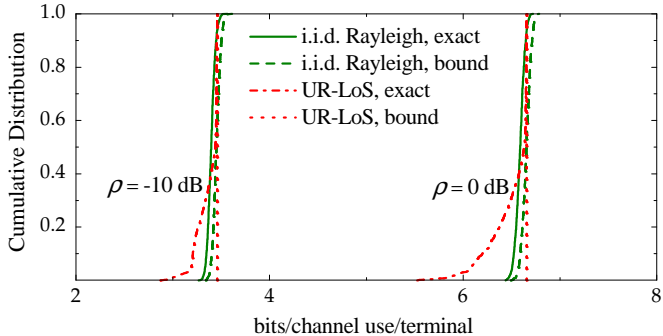}}
\caption[Spectral efficiency in i.i.d.~Rayleigh fading and
UR-LoS.]{Capacity per terminal for i.i.d.\ Rayleigh fading and
UR-LoS channels. Here  $M=100$ and $K=10$.\label{fig:fp1}}
\end{figure}

\begin{figure}[t!]
\centerline{\includegraphics[width=0.49\textwidth]{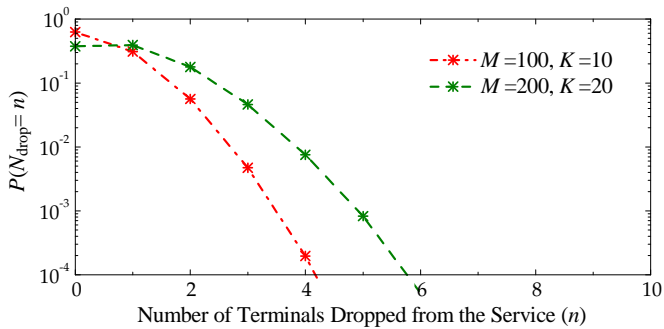}}
\caption[Probability of dropping $n$ or more terminals in
UR-LoS.]{The probability that $n$ terminals must be dropped from
service, using urns-and-balls model for UR-LoS
propagation.\label{fig:fp3_1}}\vspace{-0.5cm}
\end{figure}

\vspace{-0.5cm}

\section{Examples and Discussions}

Figure~\ref{fig:fp1} shows the cumulative probability of the
capacity per terminal for i.i.d. Rayleigh fading and UR-LoS
channels, when $M=100$ and $K=10$. The ``exact'' curves are
obtained by using (\ref{eq:macrate1}), and the ``bound'' curves
are obtained by using the upper bound (\ref{eq:Rfp1}) which  is
the maximum sum-capacity achieved under favorable propagation.
For both Rayleigh fading and UR-LoS, the sum-capacity
is very close to its upper bound with high probability. This
validates our analysis: both independent Rayleigh fading and
UR-LoS channels offer favorable propagation. Note that, despite
the fact that the condition number for UR-LoS is large with high probability
(see Fig.~\ref{fig:fp2b_1}), we only need to drop a small number of
terminals (2 terminals in this case) from  service  to have
favorable propagation. As a result, the gap between capacity and
its upper bound is very small with high probability.

  Figure~\ref{fig:fp3_1} shows the probability
that $n$ terminals must be dropped from service,
$P\left(N_{\tiny\mbox{drop}}= n \right)$, for two cases: $M=100,
K=10$ and $M=200, K=20$. This probability
 is computed by using
\eqref{eq:urlosdrop1abc}. We can see that the probability that
three terminals (for the case of $M=100$, $K=10$) and four
terminals
 (for the case of $M=200$, $K=20$) must be dropped
is less than 1\%. This is in line with  the result in
Fig.~\ref{fig:fp2b_2} where three (for the case of $M=100$,
$K=10$) or four (for the case of $M=200$, $K=20$) of the singular
values are substantially smaller than the rest, with probability
less than 1\%. Note that, to guarantee favorable propagation, the
number of terminals must be dropped is small ($\approx$ 20\%).

\vspace{-0.3cm}

\section{Conclusion} \label{Sec:Conclusion}

Both i.i.d.~Rayleigh fading and LoS with uniformly random
angles-of-arrival
 offer asymptotically favorable propagation. In i.i.d. Rayleigh, the
 channel singular values are well spread out between the smallest and largest value.
 In UR-LoS, almost all singular values are concentrated around the maximum singular value,
 and a small number of singular values are very small. Hence, in UR-LoS, by dropping a
 few terminals, the propagation is approximately favorable.

 The i.i.d.~Rayleigh and the UR-LoS
scenarios represent two extreme cases: rich scattering, and no scattering. In practice, we are
likely to have a scenario which lies in between of these two
cases. Thus, it is reasonable to expect that in most practical
environments, we have approximately favorable propagation.

The observations made regarding the UR-LoS model also underscore the importance of
performing user selection in massive MIMO.





\end{document}